\documentclass[]{article}
\usepackage{multirow}
\usepackage[bottom]{footmisc}
\usepackage[round]{natbib}

\usepackage{amsmath} 
\usepackage{amsfonts} 

\def\xixi{\mbox{\boldmath $\xi$}}
\def\pipi{\pmb{\pi}}

\def\LAM{\mbox{\boldmath $\Lambda$}}

\def\AA{\mbox{$\mathbf A$}}	\def\aa{\mbox{$\mathbf a$}}
  
\def\BB{\mbox{$\mathbf B$}}	   
\def\CC{\mbox{$\mathbf C$}}

\def\E{\,\textup{\textrm{E}}}	

\def\FF{\mbox{$\mathbf F$}} 
\def\GG{\mbox{$\mathbf G$}}	
\def\HH{\mbox{$\mathbf H$}}	
\def\II{\mbox{$\mathbf I$}}

\def\KK{\mbox{$\mathbf K$}}

\def\MVN{\,\textup{\textrm{MVN}}}

\def\QQ{\mbox{$\mathbf Q$}}	    
\def\RR{\mbox{$\mathbf R$}}	  	
\def\Ss{\mbox{$\mathbf S$}}
\def\UU{\mbox{$\mathbf U$}}	\def\uu{\mbox{$\mathbf u$}}

\def\VV{\mbox{$\pmb{V}$}}	\def\vv{\mbox{$\pmb{v}$}}
\def\WW{\mbox{$\pmb{W}$}}	\def\ww{\mbox{$\pmb{w}$}}
\def\XX{\mbox{$\pmb{X}$}}	\def\xx{\mbox{$\pmb{x}$}}
\def\YY{\mbox{$\pmb{Y}$}}	\def\yy{\mbox{$\pmb{y}$}}
\def\ZZ{\mbox{$\mathbf Z$}}	\def\zz{\mbox{$\mathbf z$}}	  

\def\chol{\,\textup{\textrm{chol}}}

\def\var{\,\textup{\textrm{var}}}
\def\cov{\,\textup{\textrm{cov}}}

\def\hatxtT{\widetilde{\xx}_t^T}
\def\hatxtpT{\widetilde{\xx}_{t+1}^T}
\def\hatxtt1{\widetilde{\xx}_t^{t-1}}

\def\hatxtmT{\widetilde{\xx}_{t-1}^T}
\def\hatxtmt1{\widetilde{\xx}_{t-1}^{t-1}}
\def\hatxtpt1{\widetilde{\xx}_{t+1}^{t-1}}

\def\hatWt{\widehat{\WW}_t}
\def\hatWtp{\widehat{\WW}_{t+1}}
\def\hatwt{\widehat{\ww}_t}
\def\hatwtp{\widehat{\ww}_{t+1}}
\def\checkWt{\overline{\WW}_{t}}
\def\checkWtp{\overline{\WW}_{t+1}}
\def\checkwt{\overline{\ww}_t}
\def\checkwtp{\overline{\ww}_{t+1}}

\def\hatVt{\widehat{\VV}_t}
\def\hatvt{\widehat{\vv}_t}
\def\checkVt{\overline{\VV}_t}
\def\checkvt{\overline{\vv}_t}
\def\hatVtT{\widetilde{\VV}_t^T}
\def\hatVtt1{\widetilde{\VV}_t^{t-1}}
\def\hatVtmT{\widetilde{\VV}_{t-1}^T}
\def\hatVtmt1{\widetilde{\VV}_{t-1}^{t-1}}
\def\hatVttmT{\widetilde{\VV}_{t,t-1}^T}
\def\hatVtmtT{\widetilde{\VV}_{t-1,t}^T}
\def\hatVttmt1{\widetilde{\VV}_{t,t-1}^{t-1}}
\def\hatVtpT{\widetilde{\VV}_{t+1}^T}
\def\hatVttpT{\widetilde{\VV}_{t,t+1}^T}
\def\hatVttpt1{\widetilde{\VV}_{t,t+1}^{t-1}}
\def\hatVtptT{\widetilde{\VV}_{t+1,t}^T}
\def\hatVtptt1{\widetilde{\VV}_{t+1,t}^{t-1}}
\def\hatUtT{\widetilde{\UU}_t^T}
\def\hatUtt1{\widetilde{\UU}_t^{t-1}}
\def\hatStT{\widetilde{\Ss}_t^T}
\def\hatStt1{\widetilde{\Ss}_t^{t-1}}
\def\hatSttmT{\widetilde{\Ss}_{t,t-1}^T}
\def\hatSttmt1{\widetilde{\Ss}_{t,t-1}^{t-1}}
\def\hatSttpT{\widetilde{\Ss}_{t,t+1}^T}
\def\hatSttpt1{\widetilde{\Ss}_{t,t+1}^{t-1}}

\usepackage[round]{natbib} 

\makeatletter
\renewcommand*\env@matrix[1][*\c@MaxMatrixCols c]{%
  \hskip -\arraycolsep
  \let\@ifnextchar\new@ifnextchar
  \array{#1}}
\makeatother
\begin{document}
\author{E. E. Holmes\footnote{Northwest Fisheries Science Center, NOAA Fisheries, Seattle, WA 98112, 
       eli.holmes@noaa.gov, http://faculty.washington.edu/eeholmes}}
\title{Computation of Standardized Residuals for MARSS Models}
\maketitle
\begin{abstract}
This report shows how to compute the variance of the joint conditional model and state residuals for multivariate autoregressive Gaussian state-space (MARSS) models. The bulk of the report focuses on `smoothations', which are the residuals conditioned on all the data $t=1$ to $T$. The final part of the report covers `innovations', which are residuals conditioned on the data $t=1$ to $t-1$.  

The MARSS model can be written: $\xx_t=\BB\xx_{t-1}+\uu+\ww_t$, $\yy_t=\ZZ\xx_t+\zz+\vv_t$, where $\ww_t$ and $\vv_t$ are independent multivariate Gaussian error-terms with variance-covariance matrices $\QQ_t$ and $\RR_t$ respectively. The joint conditional residuals are the $\ww_t$ and $\vv_t$ conditioned on the observed data, which may be incomplete (missing values). Harvey, Koopman and Penzer (1998) show a recursive algorithm for the smoothation residuals (conditioned on all the data). I show an alternate algorithm to compute these residuals using the conditional variances of the states and the conditional covariance between unobserved data and states. This allows one to compute the variance of un-observed residuals (residuals associated with missing or left-out data), which is needed for leave-one-out cross-validation tests. I show how to modify the Harvey et al. algorithm in the case of missing values and how to modify it to return the non-normalized conditional residuals.
\end{abstract}
\vfill
{\noindent \small citation: Holmes, E. E. 2014. Computation of standardized residuals for (MARSS) models. Technical Report. arXiv:1411.0045 }
 \newpage
 
\section{Overview}

This report discusses the computation of the variance of the conditional model and state residuals for MARSS models of the  form:
\begin{equation}\label{eq:residsMARSS}
\begin{gathered}
\xx_t = \BB_t\xx_{t-1} + \uu_t + \ww_t, \text{ where } \WW_t \sim \MVN(0,\QQ_t)\\
\yy_t = \ZZ_t\xx_t + \aa_t + \vv_t, \text{ where } \VV_t \sim \MVN(0,\RR_t)\\
\XX_0 \sim \MVN(\xixi,\LAM) \text{ or } \xx_0 = \pipi .
\end{gathered}
\end{equation}
The state and model residuals are respectively
\begin{equation}\label{eq:resids}
\begin{gathered}
\ww_t = \xx_t - \BB_t\xx_{t-1} - \uu_t\\
\vv_t = \yy_t - \ZZ_t\xx_t - \aa_t .
\end{gathered}
\end{equation}
The model (and state) residuals are a random variables since $\yy_t$ and $\xx_t$ are drawn from the joint multivariate distribution of $\YY_t$ and $\XX_t$ defined by the MARSS equations (Equation \ref{eq:residsMARSS}).
The unconditional\footnote{meaning not conditioning on any particular set of observed data but rather taking the expectation across all possible values of $\yy_t$ and $\xx_t$.} variance of the model residuals is
\begin{equation}\label{eq:unconditiondistofVt}
\var_{XY_t}[\VV_t] = \var_{XY_t}[\YY_t - (\ZZ_t \XX_t + \aa_t)] = \RR_t\\
\end{equation}
based on the distribution of $\VV_t$ in Equation \ref{eq:residsMARSS}.  $\var_{XY_t}$ indicates that the integration is over the joint unconditional distribution of $\XX_t$ and $\YY_t$. 

Once we have data, $\RR_t$ is not the variance-covariance matrix of our model residuals because our residuals are now conditioned\footnote{`conditioned' means that the probability distribution of the residual has changed. The distribution is now the distribution given that $\YY=\yy$, say. Expectations and variances $\var[ ]$ are integrals over the value that a random variable might take multiplied by the probability of that value. When presenting an   `expectation', the probability distribution is normally implicit but for derivations involving conditional expectations, it is important to be explicit about the distribution that is being integrated over.} on a set of observed data. There are two types of conditional model residuals used in MARSS analyses: innovations and smoothations.  Innovations are the model residuals at time $t$ using the expected value of $\XX_t$ conditioned on the data from 1 to $t-1$.  Smoothations  are the model residuals using the expected value of $\XX_t$ conditioned on all the data, $t=1$ to $T$.  Smoothations are used in computing standardized residuals for outlier and structural break detection \citep{Harveyetal1998, deJongPenzer1998, CommandeurKoopman2007}.  

\section{Distribution of MARSS smoothation residuals}\label{sec:smoothations}

This section discusses computation of the variance of the model and state residuals conditioned on all the data from $t=1$ to $T$.  These MARSS residuals are often used for outlier detection and shock detection, and in this case you only need the distribution of the model residuals for the observed values.  However if you wanted to do a leave-one-out cross-validation, you would need to know the distribution of the residuals for data points you left out (treated as unobserved).  The equations in this report give you the former and the latter, while the algorithm by \citet{Harveyetal1998} gives only the former.

\subsection{Notation and relations}

Throughout, I follow the convention that capital letters are random variables and small letters are a realization from the random variable.  This only applies to random variables; parameters are not random variables\footnote{in a frequentist framework}. Parameters are shown in Roman font while while random variables are bold slanted font. Parameters written as capital letters are matrices, while parameters written in small letters are strictly column matrices. 

In this report, the distribution over which the integration is done in an expectation or variance  is given by the subscript, e.g. $\E_A[f(A)]$ indicates an unconditional expectation over the distribution of $A$ without conditioning on another random variable while $\E_{A|b}[f(A)|b]$ would indicate an expectation over the distribution of $A$ conditioned on $B=b$; presumably $A$ and $B$ are not independent otherwise $B=b$ would have no effect on $A$. $\E_{A|b}[f(A)|b]$ is a fixed value, not random. It is the expected value when $B=b$. In contrast, $\E_{A|B}[f(A)|B]$ denotes the random variable over all the possible $\E_{A|b}[f(A)|b]$ given all the possible $b$ values that $B$ might take. The variance of $\E_{A|B}[f(A)|B]$ is the variance of this random variable. The variance of $\E_{A|b}[f(A)|b]$ in contrast is 0 since it is a fixed value.  We will often be working with the random variables, $\E_{A|B}[f(A)|B]$ or $\var_{A|B}[f(A)|B]$, inside an expectation or variance: such as $\var_B[\E_{A|B}[f(A)|B]]$.

\subsubsection{Law of total variance}

The ``law of total variance'' can be written
\begin{equation}\label{eq:lawoftotvar}
\var_A[A] = \var_B[\E_{A|B}[A|B]] + \E_B[\var_{A|B}[A|B]] .
\end{equation}
The subscripts on the inner expectations make it explicit that the expectations are being taken over the conditional distributions.  $\var_{A|B}[A|B]$ and $\E_{A|B}[A|B]$ are random variables because the $B$ in the conditional is a random variable. We take the expectation or variance with $B$ fixed at one value, $b$, but $B$ can take other values of $b$ also.  

Going forward, I will write the law or total variance more succinctly as
\begin{equation}
\var[A] = \var_B[\E[A|B]] + \E_B[\var[A|B]] .
\end{equation}
I leave off the subscript on the inner conditional expectation or variance. Just remember that when you see a conditional in an expectation or variance, the integration is over over the conditional distribution of $A$ conditioned on $B=b$. Even when you see $A|B$, the conditioning is on $B=b$ and the $B$ indicates that this is a random variable because $B$ can take different $b$ values. When computing $\var_B[\E_{A|B}[A|B]]$, we will typically compute $\E_{A|b}[A|b]$ and then compute (or infer) the variance or expectation of that over all possible values of $b$. 

The law of total variance will appear in this report in the following form:
\begin{equation}
\var_{XY_t}[f(\YY_t,\XX_t)] = \var_{Y^{(1)}}[\E_{XY_t|Y^{(1)}}[f(\YY_t,\XX_t)|\YY^{(1)}]] + \E_{Y^{(1)}}[\var_{XY_t|Y^{(1)}}[f(\YY_t,\XX_t)|\YY^{(1)}]] ,
\end{equation}
where $f(\YY_t,\XX_t)$ is some function of $\XX_t$ and $\YY_t$ and $\YY^{(1)}$ is the observed data from $t=1$ to $T$ ($\YY^{(2)}$ is the unobserved data).

\subsection{Model residuals conditioned on all the data}

Define the smoothations $\hatvt$ as:
\begin{equation}\label{eq:vtT}
\hatvt = \yy_t - \ZZ_t\hatxtT - \aa_t,
\end{equation}
where  $\hatxtT$ is $\E[\XX_t|\yy^{(1)}]$. The smoothation is different from $\vv_t$ because it uses $\hatxtT$ not $\xx_t$; $\xx_t$ is not known, and $\hatxtT$ is its estimate. $\hatxtT$ is output by the Kalman smoother. $\yy^{(1)}$ means all the observed data from $t=1$ to $T$. $\yy^{(1)}$ is a sample from the random variable $\YY^{(1)}$. The unobserved $\yy$ will be termed $\yy^{(2)}$ and is a sample from the random variable $\YY^{(2)}$. When $\YY$ appears without a superscript, it means both $\YY^{(1)}$ and $\YY^{(2)}$ together. Similarly $\yy$ means both $\yy^{(1)}$ and $\yy^{(2)}$ together---the observed data that we use to estimate $\hatxtT$ and the unobserved data that we do not use and may or may not know. $\hatvt$ exists for both $\yy^{(1)}$ and $\yy^{(2)}$, though we might not know $\yy^{(2)}$ and thus might not know its corresponding $\hatvt$. In some cases, however, we do know $\yy^{(2)}$; they are data that we left out of our model fitting, in say a k-fold or leave-one-out cross-validation.

$\hatvt$ is a sample from the random variable $\hatVt$.  We want to compute the mean and variance of this random variable over all possibles values that $\XX_t$ and $\YY_t$ might take. The mean of $\hatVt$ is 0 and we are concerned only with computing the variance:
\begin{equation}\label{eq:var.vtT}
\var[\hatVt] = \var_{XY_t}[\YY_t - \ZZ_t\E[\XX_t|\YY^{(1)}] - \aa_t] .
\end{equation}
Notice we have an unconditional variance over $XY_t$ (i.e., over all possible values that $\XX_t$ and $\YY_t$ can take) on the outside and a conditional expectation over a specific value of $\YY^{(1)}$ on the inside (in the $\E[\;]$).

From the law of total variance (Equation \ref{eq:lawoftotvar}), we can write the variance of the model residuals as
\begin{equation}\label{eq:varvvtgeneral}
\var[\hatVt] = \var_{Y^{(1)}}[\E[\hatVt|\YY^{(1)}]] + \E_{Y^{(1)}}[\var[\hatVt|\YY^{(1)}]] .
\end{equation}

\subsubsection{First term on right hand side of Equation \ref{eq:varvvtgeneral}}

The random variable inside the $\var[\;]$ in the first term is
\begin{equation}\label{eq:first.term.rhs.varvvtgeneral}
\E[\hatVt|\YY^{(1)}]= \E[(\YY_t + \ZZ_t \E[\XX_t|\YY^{(1)}] + \aa_t)|\YY^{(1)}] .
\end{equation}
Let's consider this for a specific value $\YY^{(1)}=\yy^{(1)}$.
\begin{equation}
\E[\hatVt|\yy^{(1)}] = \E[(\YY_t + \ZZ_t \E[\XX_t|\yy^{(1)}] + \aa_t)|\yy^{(1)}] =
\E[\YY_t|\yy^{(1)}] + \ZZ_t \E[\E[\XX_t|\yy^{(1)}]|\yy^{(1)}] + \E[\aa_t|\yy^{(1)}] .
\end{equation}
$\E[\XX_t|\yy^{(1)}]$ is a fixed value, and the expected value of a fixed value is itself. So $\E[\E[\XX_t|\yy^{(1)}]|\yy^{(1)}]=\E[\XX_t|\yy^{(1)}]$. 
Thus,
\begin{equation}
\E[\hatVt|\yy^{(1)}] = \E[\YY_t|\yy^{(1)}] + \ZZ_t \E[\XX_t|\yy^{(1)}] + \E[\aa_t|\yy^{(1)}] .
\end{equation}
We can move the conditional out and write
\begin{equation}
\E[\hatVt|\yy^{(1)}]= \E[(\YY_t + \ZZ_t \XX_t + \aa_t)|\yy^{(1)}]=\E[\VV_t|\yy^{(1)}].
\end{equation}
The right side is $\E[\VV_t|\yy^{(1)}]$, no `hat' on the $\VV_t$, and this applies for all $\yy^{(1)}$.  This means that the first term in Equation \ref{eq:varvvtgeneral} can be written with no hat on $\VV$:
\begin{equation}\label{eq:no.hat.on.V}
\var_{Y^{(1)}}[\E[\hatVt|\YY^{(1)}]] = \var_{Y^{(1)}}[\E[\VV_t|\YY^{(1)}]] .
\end{equation}

Using the law of total variance, we can re-write $\var[\VV_t]$ as:
\begin{equation}\label{eq:varianceVt}
\var[\VV_t] = \var_{Y^{(1)}}[\E[\VV_t|\YY^{(1)}]] + \E_{Y^{(1)}}[\var[\VV_t|\YY^{(1)}]].
\end{equation}
From Equation \ref{eq:varianceVt}, we can solve for $\var_{Y^{(1)}}[\E[\VV_t|\YY^{(1)}]]$:
\begin{equation}\label{eq:var.E.vtT}
\var_{Y^{(1)}}[\E[\VV_t|\YY^{(1)}]] = \var[\VV_t] - \E_{Y^{(1)}}[\var[\VV_t|\YY^{(1)}]].
\end{equation}
From Equation \ref{eq:unconditiondistofVt}, we know that $\var[\VV_t]=\RR_t$ (this is the unconditional variance). Thus,
\begin{equation}\label{eq:var.E.vtT.R}
\var_{Y^{(1)}}[\E[\VV_t|\YY^{(1)}]] = \RR_t - \E_{Y^{(1)}}[\var[\VV_t|\YY^{(1)}]].
\end{equation}

The second term in Equation \ref{eq:var.E.vtT.R} to the right of the equal sign and inside the expectation is $\var[\VV_t|\YY^{(1)}]$. This is the variance of $\VV_t$ with $\YY^{(1)}$ held at a specific fixed $\yy^{(1)}$. The variability in $\var[\VV_t|\yy^{(1)}]$ (notice $\yy^{(1)}$ not $\YY^{(1)}$ now) comes from $\XX_t$ and $\YY^{(2)}$ which are random variables. Let's compute this variance for a specific $\yy^{(1)}$ value.
\begin{equation}\label{eq:varvtcondy}
\var[\VV_t|\yy^{(1)}] = \var[ \YY_t - \ZZ_t\XX_t-\aa_t | \yy^{(1)} ].
\end{equation}
Notice that there is no $\E$ (expectation) on the $\XX_t$; this is $\VV_t$ not $\hatVt$. $\aa_t$ is a fixed value and can be dropped.

Equation \ref{eq:varvtcondy} can be written as\footnote{$\var(A+B)=\var(A)+\var(B)+\cov(A,B)+\cov(B,A)$}:
\begin{equation}\label{eq:var.Vt.yy}
\begin{split}
\var[\VV_t|\yy^{(1)}] &= \var[ \YY_t - \ZZ_t\XX_t | \yy^{(1)} ]\\
&=\var[ - \ZZ_t\XX_t | \yy^{(1)} ] + \var[ \YY_t|\yy^{(1)}] + \cov[ \YY_t, - \ZZ_t\XX_t | \yy^{(1)} ] + \cov[ - \ZZ_t\XX_t, \YY_t | \yy^{(1)} ]\\
&=\ZZ_t \hatVtT \ZZ_t^\top + \hatUtT - \hatStT\ZZ_t^\top - \ZZ_t(\hatStT)^\top .
\end{split}
\end{equation}
$\hatVtT = \var[ \XX_t | \yy^{(1)} ]$ and is output by the Kalman smoother. $\hatUtT=\var[\YY_t|\yy^{(1)}]$ and $\hatStT=\cov[\YY_t,\XX_t|\yy^{(1)}]$. The equations for these are given in \citet{Holmes2010} and are output by the \verb@MARSShatyt()@ function in the MARSS R package\footnote{$\hatUtT$ is  \texttt{OtT - tcrossprod(ytT)} in the \texttt{MARSShatyt()} output.}.  If there were no missing data, i.e. if $\yy^{(1)}=\yy$, then $\hatUtT$ and $\hatStT$ would be zero because $\YY_t$ would be fixed at $\yy_t$. This would reduce Equation \ref{eq:var.Vt.yy} to $\ZZ_t \hatVtT \ZZ_t^\top$. But we are concerned with the case where there are missing values. Those missing values need not be for all $t$. That is, there may be some observed $y$ at time t and some missing $y$. $\yy_t$ is multivariate.

From Equation \ref{eq:var.Vt.yy}, we know $\var[\VV_t|\yy^{(1)}]$ for a specific $\yy^{(1)}$. We want $\E_{Y^{(1)}}[\var[\VV_t|\YY^{(1)}]]$ which is its expected value over all possible values of $\yy^{(1)}$. $\hatVtT$, $\hatUtT$ and $\hatStT$ are multivariate Normal random variables. The conditional variance of a multivariate Normal does not depend on the value that you are conditioning on\footnote{Let the $\AA$ be a N-dimensional multivariate normal random variable partitioned into $\AA_1$ and $\AA_2$ with variance-covariance matrix $\Sigma = \begin{bmatrix}
\Sigma_1 & \Sigma_{12} \\
\Sigma_21 & \Sigma_{2}
\end{bmatrix}$.  The variance-covariance matrix of $\AA$ conditioned on $\AA_1=\aa$ is $\Sigma = \begin{bmatrix}
0 & 0 \\
0 & \Sigma_2 - \Sigma_{12}\Sigma_{1}\Sigma_{21}
\end{bmatrix}$. Notice that $\aa$ does not appear in the conditional variance matrix.}. This means that $\hatVtT$, $\hatUtT$ and $\hatStT$ do not depend on $\yy^{(1)}$. They only depend on the MARSS model parameters.

Because $\hatVtT$, $\hatUtT$ and $\hatStT$ only depend on the MARSS parameters values, $\QQ$, $\BB$, $\RR$, etc., the second term in Equation \ref{eq:var.E.vtT}, $\E_{Y^{(1)}}[\var[\VV_t|\YY^{(1)}]]$, is equal to $\var[\VV_t|\yy^{(1)}]$ (Equation \ref{eq:var.Vt.yy}). Putting this into Equation \ref{eq:var.E.vtT.R}, we have
\begin{equation}\label{eq:conditionalvtfinala}
\var_{Y^{(1)}}[\E[\VV_t|\YY^{(1)})]]  = \RR_t - \var[\VV_t|\yy^{(1)}] = \RR_t - \ZZ_t \hatVtT \ZZ_t^\top - \hatUtT + \hatStT\ZZ_t^\top + \ZZ_t(\hatStT)^\top.
\end{equation}
Since $\var_{Y^{(1)}}[\E[\VV_t|\YY^{(1)})]] = \var_{Y^{(1)}}[\E[\hatVt|\YY^{(1)})]]$ (Equation \ref{eq:no.hat.on.V}), this means that the first term in Equation \ref{eq:varvvtgeneral} is
\begin{equation}\label{eq:conditionalvtfinal}
\var_{Y^{(1)}}[\E[\hatVt|\YY^{(1)})]]  =  \RR_t - \ZZ_t \hatVtT \ZZ_t^\top - \hatUtT + \hatStT\ZZ_t^\top + \ZZ_t(\hatStT)^\top.
\end{equation}

\subsubsection{Second term on right hand side of Equation \ref{eq:varvvtgeneral}}

Consider the second term in Equation \ref{eq:varvvtgeneral}.  This term is 
\begin{equation}\label{eq:second.term.rhs.9}
\E_{Y^{(1)}}[\var[\hatVt|\YY^{(1)}]] = \E_{Y^{(1)}}[\var[(\YY_t-\ZZ_t\E[\XX_t|\YY^{(1)}]-\aa_t)|\YY^{(1)}]] .
\end{equation}
The middle term is:
\begin{equation}
\E_{Y^{(1)}}[\var[\E[\XX_t|\YY^{(1)}]|\YY^{(1)}]].
\end{equation}
Let's solve the inner part for a specific $\YY^{(1)}=\yy^{(1)}$. $\E[\XX_t|\yy^{(1)}]$ is a fixed value. Thus  $\var[\E[\XX_t|\yy^{(1)}]|\yy^{(1)}]=0$ since the variance of a fixed value is 0. This is true for all $\yy^{(1)}$ so the middle term reduces to 0. $\aa_t$ is also fixed and its variance is also 0. Thus for a specific $\YY^{(1)}=\yy^{(1)}$, the inside of the right hand side expectation reduces to $\var[\YY_t|\yy^{(1)}]$ which is $\hatUtT$. As noted in the previous section, $\hatUtT$ is only a function of the MARSS parameters; it is not a function of $\yy^{(1)}$ and $\var[\YY_t|\yy^{(1)}]=\hatUtT$ for all $\yy^{(1)}$. Thus the second term in Equation \ref{eq:varvvtgeneral} is simply $\hatUtT$:
\begin{equation}\label{eq:second.term.rhs.9final}
\E_{Y^{(1)}}[\var[\hatVt|\YY^{(1)}]] = \hatUtT .
\end{equation}

\subsubsection{Putting together the first and second terms}

We can now put the first and second terms in Equation \ref{eq:varvvtgeneral} together (Equations \ref{eq:conditionalvtfinal} and \ref{eq:second.term.rhs.9final}) and write out the variance of the model residuals:
\begin{equation}\label{eq:first.and.secons.vvtgeneral}
\begin{split}
\var[\hatVt] &= \RR_t - \ZZ_t \hatVtT \ZZ_t^\top - \hatUtT + \hatStT\ZZ_t^\top + \ZZ_t(\hatStT)^\top + \hatUtT\\
&= \RR_t - \ZZ_t \hatVtT \ZZ_t^\top + \hatStT\ZZ_t^\top + \ZZ_t(\hatStT)^\top .
\end{split}
\end{equation}
Equation \ref{eq:first.and.secons.vvtgeneral} will reduce to $\RR_t - \ZZ_t \hatVtT \ZZ_t^\top$ if $\yy_t$ has no missing values since $\hatStT = 0$ in this case.  If $\yy_t$ is all missing values, $\hatStT = \ZZ_t \hatVtT$ because 
\begin{equation}\label{eq:cov.Yt.Xt.no.missing.vals}
\cov[\YY_t, \XX_t|\yy^{(1)}] = \cov[\ZZ_t \XX_t + \aa_t + \VV_t, \XX_t|\yy^{(1)}] = \cov[\ZZ_t \XX_t, \XX_t|\yy^{(1)}] = \ZZ_t \cov[\XX_t, \XX_t|\yy^{(1)}] = \ZZ_t \hatVtT .
\end{equation}
The reduction in Equation \ref{eq:cov.Yt.Xt.no.missing.vals} occurs because $\VV_t$ and $\WW_t$ and by extension $\VV_t$ and $\XX_t$ are independent in the form of MARSS model used in this report (Equation \ref{eq:residsMARSS})\footnote{This is not the case for the \citet{Harveyetal1998} form of the MARSS model where $\VV_t$ and $\WW_t$ are allowed to be correlated.}. Thus when $\yy_t$ is all missing values, Equation \ref{eq:first.and.secons.vvtgeneral} will reduce to $\RR_t + \ZZ_t \hatVtT \ZZ_t^\top$ . The behavior if $\yy_t$ has some missing and some not missing values depends on whether $\RR_t$ is a diagonal matrix or not (i.e. if the $\yy_t^{(1)}$ and $\yy_t^{(2)}$ are correlated).

\subsection{State residuals conditioned on the data}

The state residuals are $\xx_t - (\BB_t \xx_{t-1} + \uu_t)=\ww_t$.  The unconditional expected value of the state residuals is $\E[\XX_t - (\BB_t \XX_{t-1} + \uu_t)] = \E[\WW_t] = 0$ and the unconditional variance of the state residuals is
\begin{equation}
\var[\XX_t - (\BB_t \XX_{t-1} + \uu_t)] = \var[\WW_t] = \QQ_t
\end{equation}
based on the definition of $\WW_t$ in Equation \ref{eq:residsMARSS}.
The conditional state residuals (conditioned on the data from $t=1$ to $t=T$) are defined as
\begin{equation}
\hatwt = \hatxtT - \BB_t\hatxtmT - \uu_t.
\end{equation}
where $\hatxtT = E[\XX_t|\yy^{(1)}]$ and $\hatxtmT = E[\XX_{t-1}|\yy^{(1)}]$.  $\hatwt$ is a sample from the random variable $\hatWt$; random over different possible data sets.  The expected value of $\hatWt$ is 0, and we are concerned with computing its variance.

We can write the variance of $\WW_t$ (no hat) using the law of total variance.
\begin{equation}\label{eq:Wlawoftotvar}
\var[\WW_t] = \var_{Y^{(1)}}[\E[\WW_t|\YY^{(1)}]] + \E_{Y^{(1)}}[\var[\WW_t|\YY^{(1)}]] .
\end{equation}
Notice that
\begin{equation}
\E[\WW_t|\yy^{(1)}] = \E[(\XX_t - \BB_t \XX_{t-1} - \uu_t)|\yy^{(1)}] =  \hatxtT - \BB_t \hatxtmT - \uu_t = \E[\hatWt|\yy^{(1)}] = \hatwt .
\end{equation}
This is true for all $\yy^{(1)}$, thus $\E[\WW_t|\YY^{(1)}]$ is $\hatWt$, and $\var_{Y^{(1)}}[\E[\WW_t|\YY^{(1)}]] = \var[\hatWt]$. Equation \ref{eq:Wlawoftotvar} can thus be written
\begin{equation}
\var[\WW_t] = \var[\hatWt] + \E_{Y^{(1)}}[\var[\WW_t|\YY^{(1)}]] .
\end{equation}
Solve for $\var[\hatWt]$:
\begin{equation}\label{eq:varwwt}
\var[\hatWt] = \var[\WW_t] - \E_{Y^{(1)}}[\var[\WW_t|\YY^{(1)}]].
\end{equation}

The variance in the expectation on the far right for a specific $\YY^{(1)}=\yy^{(1)}$ is
\begin{equation}
\var[\WW_t|\yy^{(1)}] = \var[ (\XX_t - \BB_t\XX_{t-1}-\uu_t) | \yy^{(1)} ] .
\end{equation}
$\uu_t$  is not a random variable and can be dropped. Thus\footnote{$\var[A-B]=\var[A]+\var[B]+\cov[A,-B]+\cov[-B,A]$. Be careful about the signs in this case as they are a little non-intuitive.},
\begin{equation}\label{eq:var.W.cond.y1}
\begin{split}
\var[\WW_t&|\yy^{(1)}] = \var[ (\XX_t - \BB_t\XX_{t-1}) | \yy^{(1)} ] \\
& = \var[ \XX_t | \yy^{(1)} ] + \var[\BB_t\XX_{t-1} | \yy^{(1)} ] + \cov[\XX_t, -\BB_t\XX_{t-1} | \yy^{(1)} ] + \cov[ -\BB_t\XX_{t-1}, \XX_t | \yy^{(1)} ]\\
& = \hatVtT + \BB_t \hatVtmT \BB_t^\top - \hatVttmT\BB_t^\top - \BB_t\hatVtmtT .
\end{split}
\end{equation}
Again this is conditional multivariate Normal variance, and its value does not depend on the value, $\yy^{(1)}$ that we are conditioning on.  It depends only on the parameters values, $\QQ$, $\BB$, $\RR$, etc., and is the same for all values of $\yy^{(1)}$. So $\E_{Y^{(1)}}[\var[\WW_t|\YY^{(1)}]] = \var[\WW_t|\yy^{(1)}]$, using any value of $\yy^{(1)}$. Thus
\begin{equation}\label{eq:E.var.Wt.yt}
\E_{Y^{(1)}}[\var[\WW_t|\YY^{(1)}]] =  \hatVtT + \BB_t\hatVtmT\BB_t^\top - \hatVttmT\BB_t^\top - \BB_t\hatVtmtT .
\end{equation}

Putting $\E_{Y^{(1)}}[\var[\WW_t|\YY^{(1)}]]$ from Equation \ref{eq:E.var.Wt.yt} and $\var[\WW_t]=\QQ_t$ into Equation \ref{eq:varwwt}, the variance of the conditional state residuals is
\begin{equation}
\var[\hatWt] = \QQ_t - \hatVtT - \BB_t\hatVtmT\BB_t^\top + \hatVttmT\BB_t^\top + \BB_t\hatVtmtT .
\end{equation}

\subsection{Covariance of the conditional model and state residuals}

The unconditional model and state residuals, $\VV_t$ and $\WW_t$, are independent by definition\footnote{This independence is specific to the way the MARSS model for this report (Equation \ref{eq:residsMARSS}). It is possible for the model and state residuals to covary. In the MARSS model written in \citet{Harveyetal1998} form, they do covary.} (in Equation \ref{eq:residsMARSS}), i.e., $\cov[\VV_t,\WW_t]=0$.  However the conditional model and state residuals, $\cov[\hatVt,\hatWt]$, are not independent since both depend on $\yy^{(1)}$.  
Using the law of total covariance, we can write
\begin{equation}\label{eq:covhatVtWt1}
\cov[\hatVt,\hatWt] = 
\cov_{Y^{(1)}}[\E[\hatVt|\YY^{(1)}],\E[\hatWt|\YY^{(1)}]] + \E_{Y^{(1)}}[\cov[\hatVt, \hatWt|\YY^{(1)}]] .
\end{equation}

For a specific value of $\YY^{(1)}=\yy^{(1)}$, the covariance in the second term on the right is $\cov[\hatVt, \hatWt|\yy^{(1)}]$. Conditioned on a specific value of $\YY^{(1)}$, $\hatWt$ is a fixed value, $\hatwt = \hatxtT - \BB_t\hatxtmT - \uu_t$, and conditioned on $\yy^{(1)}$, $\hatxtT$ and $\hatxtmT$ are fixed values. $\uu_t$ is also fixed; it is a parameter. $\hatVt$ is not a fixed value because it has $\YY_t^{(2)}$ and that is a random variable.  Thus $\cov[\hatVt, \hatWt|\yy^{(1)}]$ is the covariance between a random variable and a fixed variable and thus the covariance is 0. This is true for all $\yy^{(1)}$.
Thus the second right-side term in Equation \ref{eq:covhatVtWt1} is zero, and the equation reduces to
\begin{equation}\label{eq:covhatVtWt3}
\cov[\hatVt,\hatWt] = \cov_{Y^{(1)}}[\E[\hatVt|\YY^{(1)}],\E[\hatWt|\YY^{(1)}]].
\end{equation}
Notice that $\E[\hatWt|\yy^{(1)}]=\E[\WW_t|\yy^{(1)}]$ and $\E[\hatVt|\yy^{(1)}]=\E[\VV_t|\yy^{(1)}]$ since
\begin{equation}
\E[\WW_t|\yy^{(1)}]= \E[\XX_t|\yy^{(1)}]-\BB_t\E[\XX_{t-1}|\yy^{(1)}] - \uu_t = \hatxtT-\BB_t\hatxtmT - \uu_t = \hatwt = \E[\hatWt|\yy^{(1)}]
\end{equation}
and
\begin{equation}
\E[\VV_t|\yy^{(1)}]= \E[\YY_t|\yy^{(1)}]-\ZZ_t\E[\XX_{t}|\yy^{(1)}] - \aa_t = \E[\YY_t|\yy^{(1)}]-\ZZ_t \hatxtT - \aa_t = \E[\hatVt|\yy^{(1)}] .
\end{equation}
Thus the right side of Equation \ref{eq:covhatVtWt3} can be written in terms of $\VV_t$ and $\WW_t$ instead of $\hatVt$ and $\hatWt$:
\begin{equation}\label{eq:covhatVtWt2}
\cov[\hatVt,\hatWt] = \cov_{Y^{(1)}}[\E[\VV_t|\YY^{(1)}],\E[\WW_t|\YY^{(1)}]].
\end{equation}

Using the law of total covariance, we can write:
\begin{equation}\label{eq:covVtWt}
\cov[\VV_t, \WW_t] = \E_{Y^{(1)}}[\cov[\VV_t, \WW_t|\YY^{(1)}]] + \cov_{Y^{(1)}}[\E[\VV_t|\yy^{(1)}],\E[\WW_t|\YY^{(1)}]] .
\end{equation}
The unconditional covariance of $\VV_t$ and $\WW_t$  is 0. Thus the left side of Equation \ref{eq:covVtWt} is 0 and we can rearrange the equation as
\begin{equation}\label{eq:covVtWt2}
\cov_{Y^{(1)}}[\E[\VV_t|\YY^{(1)}],\E[\WW_t|\YY^{(1)}]] = - \E_{Y^{(1)}}[\cov[\VV_t, \WW_t|\YY^{(1)}]] .
\end{equation}
Combining Equation \ref{eq:covhatVtWt2} and \ref{eq:covVtWt2}, we get
\begin{equation}\label{eq:conditionalcovvtwt}
\cov[\hatVt,\hatWt] = - \E_{Y^{(1)}}[ \cov[\VV_t, \WW_t|\YY^{(1)}] ] ,
\end{equation}
and our problem reduces to solving for the conditional covariance of the model and state residuals (right side of Equation \ref{eq:conditionalcovvtwt}).  

For a specific $\YY^{(1)}=\yy^{(1)}$, the conditional covariance $\cov[\VV_t, \WW_t|\yy^{(1)}]$ can be written out as
\begin{equation}
\cov[\VV_t, \WW_t|\yy^{(1)}] = \cov[\YY_t-\ZZ_t\XX_t-\aa_t,\, \XX_t-\BB_t\XX_{t-1}-\uu_t|\yy^{(1)}] .
\end{equation}
$\aa_t$ and $\uu_t$ are fixed values and can be dropped. Thus\footnote{$\cov[\BB \mathbf{A},\CC \mathbf{D}]=\BB\cov[\mathbf{A},\mathbf{D}]\CC^\top$.}
\begin{equation}
\begin{split}
\cov&[\VV_t, \WW_t|\yy^{(1)}] =\cov[\YY_t-\ZZ_t\XX_t, \XX_t-\BB_t\XX_{t-1}|\yy^{(1)}] \\
& =\cov[\YY_t,\XX_t|\yy^{(1)}] + \cov[\YY_t,-\BB_t\XX_{t-1}|\yy^{(1)}] + \cov[-\ZZ_t\XX_t,\XX_t|\yy^{(1)}] + \cov[-\ZZ_t\XX_t,-\BB_t\XX_{t-1}|\yy^{(1)}]\\
& = \hatStT - \hatSttmT\BB_t^\top - \ZZ_t \hatVtT + \ZZ_t\hatVttmT\BB_t^\top ,
\end{split}
\end{equation}
where $\hatStT=\cov[\YY_t,\XX_t|\yy^{(1)}]$ and $\hatSttmT=\cov[\YY_t,\XX_{t-1}|\yy^{(1)}]$; the equations for $\hatStT$ and $\hatSttmT$ are given in \citet{Holmes2010} and are output by the \verb@MARSShatyt()@ function in the MARSS R package.

$\hatVtT$, $\hatVttmT$, $\hatStT$ and $\hatSttmT$ are conditional multivariate Normal and are only functions of the MARSS parameters not of $\yy^{(1)}$. Thus 
\begin{equation}
\E_{Y^{(1)}}[ \cov[\VV_t, \WW_t|\YY^{(1)}] ]= \cov[\VV_t, \WW_t|\yy^{(1)}] = \hatStT - \hatSttmT\BB_t^\top - \ZZ_t \hatVtT + \ZZ_t\hatVttmT\BB_t^\top .
\end{equation}
$\cov[\hatVt,\hatWt]$ is the negative of this (Equation \ref{eq:conditionalcovvtwt}), thus
\begin{equation}
\cov[\hatVt,\hatWt] = - \hatStT + \hatSttmT\BB_t^\top + \ZZ_t \hatVtT - \ZZ_t\hatVttmT\BB_t^\top .
\end{equation}

The Harvey et al. algorithm (next section) gives the joint distribution of the model residuals at time $t$ and state residuals at time $t+1$.  Using the law of total covariance as above, the covariance in this case is
\begin{equation}
\cov_{Y^{(1)}}[\E[\VV_t|\YY^{(1)}],\E[\WW_{t+1}|\YY^{(1)}]] = - \E_{Y^{(1)}}[ \cov[\VV_t, \WW_{t+1}|\YY^{(1)}] ] 
\end{equation}
and
\begin{equation}
\begin{split}
\cov[\VV_t, \WW_{t+1}|\yy^{(1)}] & =\cov[\YY_t-\ZZ_t\XX_t-\aa_t,\, \XX_{t+1}-\BB_{t+1}\XX_t-\uu_{t+1}|\yy^{(1)}] \\
& =\cov[\YY_t-\ZZ_t\XX_t,\, \XX_{t+1}-\BB_{t+1}\XX_t|\yy^{(1)}] \\
& = \hatSttpT - \hatStT\BB_{t+1}^\top - \ZZ_t\hatVttpT + \ZZ_t \hatVtT \BB_{t+1}^\top .
\end{split}
\end{equation}
Thus,
\begin{equation}
\begin{split}
\cov_{Y^{(1)}}[\E[\VV_t|\YY^{(1)}],\E[\WW_{t+1}|\YY^{(1)}]] & = - \E_{Y^{(1)}}[ \cov[\VV_t, \WW_{t+1}|\YY^{(1)}] ] \\ 
& = - \hatSttpT + \hatStT\BB_{t+1}^\top + \ZZ_t\hatVttpT - \ZZ_t\hatVtT\BB_{t+1}^\top.
\end{split}
\end{equation}

\subsection{Joint distribution of the conditional residuals}

We now can write the variance of the joint distribution of the conditional residuals. Define
\begin{equation}
\widehat{\varepsilon}_t = \begin{bmatrix}\hatvt\\ \hatwt\end{bmatrix} =
\begin{bmatrix}\yy_t - \ZZ_t\hatxtT - \aa_t\\ \hatxtT - \BB_t\hatxtmT - \uu_t \end{bmatrix}.
\end{equation}
$\widehat{\varepsilon}_t$ is a sample drawn from the distribution of the random variable $\widehat{\mathcal{E}}_t$.  The expected value of $\widehat{\mathcal{E}}_t$ over all possible $\yy$ is 0 and the variance of $\widehat{\mathcal{E}}_t$  is
\begin{equation}
 \begin{bmatrix}[c|c]
 \var[\hatVt]&
 \cov[\hatVt, \hatWt] \\
 \rule[.5ex]{20ex}{0.25pt} & \rule[.5ex]{20ex}{0.25pt} \\
 (\cov[\hatVt, \hatWt])^\top& 
 \var[\hatWt] \end{bmatrix}
\end{equation}
which is
\begin{equation}\label{eq:jointcondresid1general}
 \begin{bmatrix}[c|c]
 \RR_t - \ZZ_t\hatVtT\ZZ_t^\top + \hatStT\ZZ_t^\top+\ZZ_t(\hatStT)^\top&
 - \hatStT + \hatSttmT\BB_t^\top + \ZZ_t\hatVtT - \ZZ_t\hatVttmT\BB_t^\top\\
 \rule[.5ex]{40ex}{0.25pt} & \rule[.5ex]{50ex}{0.25pt} \\
 (- \hatStT + \hatSttmT\BB_t^\top + \ZZ_t\hatVtT  - \ZZ_t\hatVttmT\BB_t^\top)^\top& 
 \QQ_t - \hatVtT - \BB_t\hatVtmT\BB_t^\top + \hatVttmT\BB_t^\top + \BB_t\hatVtmtT \end{bmatrix} .
\end{equation}

If the residuals are defined as in \citet{Harveyetal1998},
\begin{equation}
\widehat{\varepsilon}_t = \begin{bmatrix}\hatvt\\ \hatwtp \end{bmatrix} =
\begin{bmatrix}\yy_t - \ZZ_t\hatxtT - \aa_t\\ \hatxtpT - \BB_{t+1}\hatxtT - \uu_{t+1} \end{bmatrix}
\end{equation}
and the variance of $\widehat{\mathcal{E}}_t$ is
\begin{equation}
 \begin{bmatrix}[c|c]
 \var[\hatVt]&
 \cov[\hatVt, \hatWtp] \\
 \rule[.5ex]{20ex}{0.25pt} & \rule[.5ex]{20ex}{0.25pt} \\
 (\cov[\hatVt, \hatWtp])^\top& 
 \var[\hatWtp] \end{bmatrix}
\end{equation}
which is
\begin{equation}\label{eq:jointcondresid2}
\begin{bmatrix}[c|c]
\RR_t - \ZZ_t\hatVtT\ZZ_t^\top + \hatStT\ZZ_t^\top + \ZZ_t(\hatStT)^\top&
- \hatSttpT + \hatStT\BB_{t+1}^\top + \ZZ_t\hatVttpT - \ZZ_t\hatVtT\BB_{t+1}^\top \\
\rule[.5ex]{40ex}{0.25pt} & \rule[.5ex]{50ex}{0.25pt} \\
(- \hatSttpT + \hatStT\BB_{t+1}^\top + \ZZ_t\hatVttpT - \ZZ_t\hatVtT\BB_{t+1}^\top)^\top& 
\QQ_{t+1} - \hatVtpT - \BB_{t+1}\hatVtT\BB_{t+1}^\top + \hatVtptT \BB_{t+1}^\top + \BB_{t+1}\hatVttpT \end{bmatrix} .
\end{equation}

The above gives the variance of both `observed' model residuals (the ones associated with $\yy^{(1)}$) and the unobserved model residuals (the ones associated with $\yy^{(2)}$).  
When there are no missing values in $\yy_t$, the $\hatStT$ and $\hatSttmT$ terms equal 0 and drop out.

\section{Harvey et al. 1998 algorithm for the conditional residuals}
\citet[pgs 112-113]{Harveyetal1998} give a recursive algorithm for computing the variance of the conditional residuals when the time-varying MARSS equation is written as: 
\begin{equation}\label{eq:residsMARSSHarvey}
\begin{gathered}
\xx_{t+1} = \BB_{t+1}\xx_t + \uu_{t+1} + \HH_{t+1}\epsilon_{t},\\
\yy_t = \ZZ_t\xx_t + \aa_t + \GG_t\epsilon_t,\\
\mbox{ where } \epsilon_t \sim \MVN(0,\II_{m+n \times m+n}) \\
\HH_t\HH_t^\top=\QQ_t, \GG_t\GG_t^\top=\RR_t, \text{ and } \HH_t\GG_t^\top = \cov[\WW_t, \VV_t]
\end{gathered}
\end{equation}
The $\HH_t$ and $\GG_t$ matrices specify the variance and covariance of $\WW_t$ and $\VV_t$. $\HH_t$ has $m$  rows and $m+n$ columns and $\GG_t$ has $n$ rows and $m+n$ columns. In the MARSS equation for this report (Equation \ref{eq:residsMARSS}), $\WW_t$ and $\VV_t$ are independent. To achieve this in the Harvey et al. form (Equation \ref{eq:residsMARSSHarvey}), the first $n$ columns of $\HH_t$ are all 0 and the last $m$ columns of $\GG_t$ are all zero.  

The algorithm in \citet{Harveyetal1998} gives the variance of the `normalized' residuals, the $\epsilon_t$.  I have modified their algorithm so it  returns the `non-normalized' residuals:
$$\varepsilon_t=\begin{bmatrix}\GG_t\epsilon_t\\ \HH_{t+1}\epsilon_t\end{bmatrix}=\begin{bmatrix}\vv_t\\ \ww_{t+1} \end{bmatrix}.$$

The Harvey et al. algorithm is a backwards recursion using the following output from the Kalman filter: the one-step ahead prediction covariance $\FF_t$, the Kalman gain $\KK_t$, $\hatxtt1=\E[\XX_t|\yy^{(1),1:{t-1}}]$ and $\hatVtt1=\var[\XX_t|\yy^{(1),1:{t-1}}]$. In the MARSS R package, these are output from \texttt{MARSSkfss()} in \texttt{Sigma}, \texttt{Kt}, \texttt{xtt1} and \texttt{Vtt1}.

\subsection{Algorithm}

Start from $t=T$ and work backwards to $t=1$. At time $T$, $r_T=0_{1 \times m}$ and $N_T=0_{m \times m}$. $\BB_{t+1}$ and $\QQ_{t+1}$ can be set to NA or 0. They will not appear in the algorithm at time $T$ since $r_T=0$ and $N_T=0$. Note that the $\ww$ residual and its associated variance and covariance with $\vv$ at time $T$ is NA since this residual would be for $\xx_T$ to $\xx_{T+1}$.


\begin{equation}\label{eq:Harveyalgo}
\begin{gathered}
\QQ^\prime_{t+1}=\begin{bmatrix}0_{m \times n}&\QQ_{t+1}\end{bmatrix}, \mbox{    } \RR^\prime_t=\begin{bmatrix}\RR_t^* & 0_{n \times m}\end{bmatrix}\\
\FF_t = \ZZ_t^*\hatVtt1{\ZZ_t^*}^\top+\RR_t^* , \,\, n \times n \\
K_t = \BB_{t+1}\KK_t = \BB_{t+1} \hatVtt1{\ZZ_t^*}^\top \FF_t^{-1}  , \,\, m \times n  \\
L_t = \BB_{t+1} - K_t\ZZ_t^*  , \,\, m \times m \\
J_t= \QQ^\prime_{t+1} - K_t \RR^\prime_t  , \,\, m \times (n+m) \\
v_t = \yy_t^* - \ZZ_t\hatxtt1 - \aa_t , \,\, n \times 1 \\
u_t = \FF_t^{-1} v_t - K_t^\top r_t , \,\, n \times 1 \\
r_{t-1} = {\ZZ_t^*}^\top u_t + \BB_{t+1}^\top r_t , \,\, m \times 1  \\
N_{t-1} = {\ZZ_t^*}^\top \FF_t^{-1} \ZZ_t^* + L_t^\top N_t L_t   , \,\, m \times m .
\end{gathered}
\end{equation}
$\yy_t^*$ is the observed data at time $t$ with the $i$-th rows set to 0 if the $i$-th $y$ is missing. 
Bolded terms are the same as in Equation \ref{eq:residsMARSSHarvey} (and are output by \texttt{MARSSkfss()}).  Unbolded terms are terms used in \citet{Harveyetal1998}.  The * on $\ZZ_t$ and $\RR_t$, indicates that they are the missing value modified versions  discussed in \citet[section 6.4]{ShumwayStoffer2006} and \citet{Holmes2010}: to construct $\ZZ_t^*$ and $\RR_t^*$, the rows of $\ZZ_t$ corresponding to missing rows of $\yy_t$ are set to zero and the $(i,j)$ and $(j,i)$ terms of $\RR_t$ corresponding the missing rows of $\yy_t$ are set to zero.  For the latter, this means if the $i$-th row of $\yy_t$ is missing, then then the $i$-th row and column (including the value on the diagonal) in $\RR_t$ are set to 0. Notice that $\FF_t$ will have 0's on the diagonal if there are missing values. A modified inverse of $\FF_t$ is used: any 0's on the diagonal of $\FF_t$ are replaced with 1, the inverse is taken, and 1s on diagonals is replaced back with 0s.

The residuals \citep[eqn 24]{Harveyetal1998} are
\begin{equation}\label{eq:Harveyresiduals}
\widehat{\varepsilon}_t = \begin{bmatrix}\hatvt\\ \hatwtp \end{bmatrix} =({\RR^\prime_t})^\top u_t + ({\QQ^\prime_{t+1}})^\top r_t
\end{equation}
The expected value of $\widehat{\mathcal{E}}_t$ is 0 and its variance is
\begin{equation}\label{eq:Harveyvariance}
\Sigma_t = \var_{Y^{(1)}}[\widehat{\mathcal{E}}_t] ={\RR^\prime_t}^\top \FF_t^{-1} \RR^\prime_t + J_t^\top N_t J_t .
\end{equation}
These $\widehat{\varepsilon}_t$ and $\Sigma_t$ are for both the non-missing and missing $\yy_t$. This is a modification to the \citet{Harveyetal1998} algorithm which does not give the variance for missing $\yy$.

\subsection{Difference in notation}

In Equation 20 in \citet{Harveyetal1998}, their $T_t$ is my $\BB_{t+1}$ and their $H_t H_t^\top$ is my $\QQ_{t+1}$.  Notice the difference in the time indexing. My time indexing on $\BB$ and $\QQ$ matches the left $\xx$ while in theirs, $T$ and $H$ indexing matches the right $\xx$. Thus in my implementation of their algorithm \citep[eqns. 21-24]{Harveyetal1998}, $\BB_{t+1}$ appears in place of $T_t$ and $\QQ_{t+1}$ appears in place of $H_t$. See comments below on normalization and the difference between $\QQ$ and $H$. 

\citet[eqns. 19, 20]{Harveyetal1998} use $G_t$ to refer to the $\chol(\RR_t)^\top$ (non-zero part of the $n \times n+m$ matrix) and $H_t$ to refer to $\chol(\QQ_t)^\top$.  I have replaced these with $\RR_t^\prime$ and $\QQ_t^\prime$ (Equation \ref{eq:Harveyalgo}) which causes my variant of their algorithm (Equation \ref{eq:Harveyalgo}) to give the `non-normalized' variance of the residuals. The residuals function in the MARSS R package has an option to give either normalized or non-normalized residuals.

$\KK_t$ is the Kalman gain output by the MARSS R package \verb@MARSSkf()@ function.  The Kalman gain as used in the \citet{Harveyetal1998} algorithm is $K_t=\BB_{t+1}\KK_t$. Notice that Equation 21 in \citet{Harveyetal1998} has $H_t G_t^\top$ in the equation for $K_t$. This is the covariance of the state and observation errors, which is allowed to be non-zero given the way Harvey et al. write the errors in their Equations 19 and 20. The way the MARSS R package model is written, the state and observation errors are independent of each other. Thus $H_t G_t^\top = 0$ and this term drops out of the $K_t$ equation in Equation \ref{eq:Harveyalgo}.

\subsection{Computing the standardized residuals}

The standardized residuals are computed by multiplying $\widehat{\varepsilon}_t$ by the inverse of the square root\footnote{Not the element-wise square-root. This means take the Cholesky decomposition of $\Sigma$ and then the inverse of that.} of the variance-covariance matrix for $\widehat{\varepsilon}_t$:
\begin{equation}
(\Sigma_t)^{-1/2}\widehat{\varepsilon}_t .
\end{equation}

\section{Distribution of the MARSS innovation residuals}

One-step-ahead predictions (innovations) are often shown for MARSS models and these are used for likelihood calculations. Innovations are the difference between the data at time $t$ minus the prediction of $\yy_t$ given data up to $t-1$. This section gives the residual variance for the innovations and the analogous values for the states.

\subsubsection{Variance of the one-step-ahead model residuals}

Define the innovations $\checkvt$ as\footnote{This is slightly different than the `innovations' that we normally use. Normally, we would work with the observed innovations while now we are talking about a sample from the random variable `innovations' not the specific sample that we observe. $\yy_t$ here is not the actual data that you observe. It's the data that you could observe. $\yy_t$ is a sample from the random variable $\YY_t$.}:
\begin{equation}\label{eq:vtt1}
\checkvt = \yy_t - \ZZ_t\hatxtt1 - \aa_t,
\end{equation}
where $\hatxtt1$ is $\E[\XX_t|\yy^{(1),t-1}]$ (expected value of $\XX_t$ conditioned on the data up to time $t-1$). The random variable, innovations over all possible $\yy_t$, is $\checkVt$. Its mean is 0 and we want to find its variance.

The derivation of the variance of $\checkVt$ follows the exact same steps as the smoothations $\hatVt$, and we can write the variance as:
\begin{equation}\label{eq:innov.model}
\var[\checkVt] = \RR_t - \ZZ_t \hatVtt1 \ZZ_t^\top + \hatStt1\ZZ_t^\top + \ZZ_t(\hatStt1)^\top
\end{equation}
where the $\hatVtt1$ and $\hatStt1$ are now conditioned on only the data from 1 to $t-1$.
$\hatStt1 = \cov[\YY_t,\XX_t|\yy^{(1), t-1}] = \cov[\ZZ_t \XX_t + \aa_t+\VV_t,\XX_t|\yy^{(1), t-1}]$. 
$\yy_t$ is not in the conditional since it only includes data up to $t-1$. Without $\yy_t$ in the conditional, $\VV_t$ and $\WW_t$ and by extension $\VV_t$ and $\XX_t$ are independent\footnote{Given the way the MARSS equation is written in this report. This is not the case for the more general Harvey et al. MARSS model which allows covariance.} and $\cov[\ZZ_t \XX_t + \aa_t + \VV_t,\XX_t|\yy^{(1), t-1}] = \cov[\ZZ_t \XX_t,\XX_t|\yy^{(1), t-1}] = \ZZ_t \hatVtt1$. Therefore, $\ZZ_t(\hatStt1)^\top =  \ZZ_t \hatVtt1 \ZZ_t^\top = \hatStt1(\ZZ_t)^\top$. Thus Equation \ref{eq:innov.model} reduces to
\begin{equation}\label{eq:innov.model2}
\var[\checkVt] = \RR_t + \ZZ_t \hatVtt1 \ZZ_t^\top.
\end{equation}

\subsection{State residuals conditioned on the data}

Define the state residuals conditioned on the data from 1 to $t-1$ as $\checkwt$.
\begin{equation}\label{eq:wtt1}
\checkwt = \hatxtt1 - \BB_t\hatxtmt1 - \uu_t,
\end{equation}
where $\hatxtmt1$ is the $\E[\XX_{t-1}|\yy^{(1),t-1}]$ (expected value of $\XX_{t-1}$ conditioned on the data up to time $t-1$). However $\hatxtt1 = \E[\XX_t|\yy^{(1),t-1}] = \BB_t\hatxtmt1 + \uu_t$. Thus $\checkwt = 0$.

\subsection{Covariance of the conditional model and state residuals}

Since $\checkwt = 0$ for all $\yy^{(1)}$:

\begin{equation}
\cov[\checkVt, \checkWt] = 0 \,\, \text{and} \,\, \cov[\checkVt, \checkWtp] = 0 .
\end{equation}

\subsection{Joint distribution of the conditional residuals}

We now the write the variance of the joint distribution of the conditional one-step ahead residuals. Define
\begin{equation}
\overline{\varepsilon}_t = \begin{bmatrix}\checkvt\\ \checkwt\end{bmatrix} =
\begin{bmatrix}\yy_t - \ZZ_t\hatxtt1-\aa_t\\ \hatxtt1 - \BB_t\hatxtmt1 - \uu_t \end{bmatrix}.
\end{equation}
where $\hatxtt1$ and $\hatxtmt1$ are conditioned on the observed $\yy$ from $t=1$ to $t-1$.
The expected value of $\overline{\mathcal{E}}_t$ over all possible $\yy$ up to time $t-1$ is 0 and the variance of $\overline{\mathcal{E}}_t$  is
\begin{equation}\label{eq:jointcondresid1innovs}
 \begin{bmatrix}[c|c]
 \var[\checkVt]&
 \cov[\checkVt, \checkWt] \\
 \rule[.5ex]{15ex}{0.25pt} & \rule[.5ex]{15ex}{0.25pt} \\
 (\cov[\checkVt, \checkWt])^\top& 
 \var[\checkWt] \end{bmatrix} =
 \begin{bmatrix}[c|c]
 \RR_t + \ZZ_t \hatVtt1 \ZZ_t^\top& 0 \\
 \rule[.5ex]{15ex}{0.25pt} & \rule[.5ex]{15ex}{0.25pt} \\
 0 & 0 \end{bmatrix} .
\end{equation}
If the residuals are defined as in \citet{Harveyetal1998},
\begin{equation}
\overline{\varepsilon}_t = \begin{bmatrix}\checkvt\\ \checkwtp\end{bmatrix} =
\begin{bmatrix}\yy_t - \ZZ_t\hatxtt1-\aa_t\\ \hatxtpt1 - \BB_{t+1}\hatxtt1-\uu_{t+1} \end{bmatrix}
\end{equation}
since $\checkWtp = 0$, and the variance-covariance matrix is again Equation \ref{eq:jointcondresid1innovs}.

\bibliography{./EMDerivation}

\begin{thebibliography}{}

\bibitem[Commandeur and Koopman, 2007]{CommandeurKoopman2007}
Commandeur, J.~J. and Koopman, S.~J. (2007).
\newblock {\em An introduction to state space time series analysis}.
\newblock Practical Econometrics. Oxford University Press, Oxford.

\bibitem[de~Jong and Penzer, 1998]{deJongPenzer1998}
de~Jong, P. and Penzer, J. (1998).
\newblock Diagnosing shocks in time series.
\newblock {\em Journal of the American Statistical Association},
  93(442):796--806.

\bibitem[Harvey et~al., 1998]{Harveyetal1998}
Harvey, A., Koopman, S.~J., and Penzer, J. (1998).
\newblock Messy time series: a unified approach.
\newblock {\em Advances in Econometrics}, 13:103--143.

\bibitem[Holmes, 2012]{Holmes2010}
Holmes, E.~E. (2012).
\newblock Derivation of the {EM} algorithm for constrained and unconstrained
  {MARSS} models.
\newblock Technical report, arXiv:1302.3919 [stat.ME].

\bibitem[Shumway and Stoffer, 2006]{ShumwayStoffer2006}
Shumway, R. and Stoffer, D. (2006).
\newblock {\em Time series analysis and its applications}.
\newblock Springer-Science+Business Media, LLC, New York, New York, 2nd
  edition.

\end{thebibliography}
\bibliographystyle{apalike}

\end{document}